\documentclass[journal]{IEEEtran}
\usepackage{graphicx,epsfig,amsfonts,amsmath,amssymb,url,ifthen}
\usepackage{times}
\usepackage{bm}
\usepackage{multirow}
\usepackage{subfigure}
\usepackage{algorithm}
\usepackage{algorithmic}

\usepackage{amsopn}
\usepackage{cases}
\usepackage{url}
\usepackage{epsfig}
\usepackage{graphicx,subfig}
\usepackage{graphicx}
\usepackage{subfigure}
\usepackage{cite}
\usepackage{amsthm}
\usepackage{CJK}
\usepackage{color}
\usepackage{makecell}
\usepackage{enumitem}
\usepackage{multirow,tabularx}
\graphicspath{{./Figures/}}

\newcommand{\tx}{\bm{x}}
\newcommand{\ty}{\bm{y}}
\newcommand{\tz}{\bm{z}}

\newcommand{\tH}{\textbf{H}}
\newcommand{\tD}{\textbf{D}}

\newcommand{\tA}{\textbf{A}}
\newcommand{\tB}{\textbf{B}}
\newcommand{\tI}{\textbf{I}}
\newcommand{\tZ}{\textbf{Z}}

\newcommand{\tR}{\textbf{R}}
\newcommand{\tW}{\textbf{W}}

\newcommand{\tTheta}{\mathbf{\Theta}}

\newcommand{\tn}{\bm{n}}

\newcommand{\tr}{\bm{r}}

\newcommand{\tw}{\bm{w}}

\newcommand{\valpha}{\bm{\alpha}}

\newcommand{\vmu}{\bm{\mu}}

\DeclareMathOperator*{\argmin}{argmin}

\title{Learning Hybrid Sparsity Prior for Image Restoration: Where Deep Learning Meets Sparse Coding}

\author{ Fangfang~Wu, Weisheng~Dong, ~\IEEEmembership{Member,~IEEE}, Guangming Shi, ~\IEEEmembership{Senior member,~IEEE}, and Xin Li, ~\IEEEmembership{Senior Member,~IEEE}
\thanks{The work was supported by the Natural Science Foundation of China under Grants (No. 61622210, 61471281, 61632019, 61472301, and 61390512, 61372131).}
\thanks{Fangfang Wu is with School of Electronic Engineering, Xidian University, Xi’an 710071, China (e-mail: ffwu1116@163.com).}
\thanks{W. Dong is with School of Artificial Intelligence, Xidian University, Xi’an 710071, China (e-mail: wsdong@mail.xidian.edu.cn).}
\thanks{G. Shi is with School of Electronic Engineering, Xidian University, Xi’an 710071, China (e-mail: gmshi@xidian.edu.cn)}
\thanks{X. Li is with the Lane Dep. of CSEE, West Virginia University, Morgantown, WV 26506-6109 USA. }}

\newboolean{doublecolumn}
\setboolean{doublecolumn}{true}

\begin{document}
\maketitle

\ifthenelse {\boolean{doublecolumn}} {} {\baselineskip=.8cm}

\begin{abstract}
State-of-the-art approaches toward image restoration can be classified into model-based and learning-based. The former - best represented by sparse coding techniques - strive to exploit intrinsic prior knowledge about the unknown high-resolution images; while the latter - popularized by recently developed deep learning techniques - leverage external image prior from some training dataset. It is natural to explore their middle ground and pursue a hybrid image prior capable of achieving the best in both worlds. In this paper, we propose a systematic approach of achieving this goal called Structured Analysis Sparse Coding (SASC). Specifically, a structured sparse prior is learned from extrinsic training data via a deep convolutional neural network (in a similar way to previous learning-based approaches); meantime another structured sparse prior is internally estimated from the input observation image (similar to previous model-based approaches). Two structured sparse priors will then be combined to produce a hybrid prior incorporating the knowledge from both domains. To manage the computational complexity, we have developed a novel framework of implementing hybrid structured sparse coding processes by deep convolutional neural networks. Experimental results show that the proposed hybrid image restoration method performs comparably with and often better than the current state-of-the-art techniques.
\end{abstract}

\begin{IEEEkeywords}
deep convolutional neural networks, structured analysis sparse coding, hybrid prior learning, image restoration.
\end{IEEEkeywords}

\IEEEpeerreviewmaketitle

\section{Introduction}
Image restoration refers to a class of ill-posed inverse problems recovering unknown images from their degraded observations (e.g., noisy, blurred or down-sampled). It is well known image prior (a.k.a. regularization) plays an important role in the development of solution algorithms to ill-posed image restoration problems. Depending on the availability of training data, one can obtain image prior by either model-based or learning-based approaches. In model-based approaches, image prior is obtained by mathematical construction of a penalty functional (e.g., total-variation or sparse coding) and its parameters have to be \emph{intrinsically} estimated from the observation data; in learning-based approaches, image prior is leveraged \emph{externally} from training data - e.g., a deep convolutional neural network is trained to learn the mapping from the space of degraded images to that of restored ones. We will briefly review the key advances within each paradigm in the past decade, which serves as the motivation for developing a hybrid (internal+external) prior in this work.

In model-based approaches, sparse coding and its variations are likely to be the most studied in the literature \cite{yu2010image,marquina2008image,yang2010image,yu2012solving,dong2011image,dong2011centralized,dong2013nonlocally,timofte2014a+,timofte2016seven,dong2016image,osendorfer2014image,wang2015deep,kim2016accurate,kim2016deeply,egiazarian2015single}.
The basic idea behind sparse coding is that natural images admit sparse representations in a transformed space. Early works in sparse coding have focused on the characterization of localized structures or transient events in natural images; to obtain basis functions with good localization properties in both spatial and frequency domains, one can either construct them through mathematical design (e.g., wavelet \cite{mallat1999wavelet}) or learn them from training data (e.g., dictionary learning \cite{mairal2009online}). Later on the importance of exploiting nonlocal similarity in natural images (e.g., self-repeating patterns in textured regions) was recognized in a flurry of so-called simultaneous sparse coding works including BM3D \cite{dabov2007image} and LSSC \cite{mairal2009non} as well as nonlocal sparsity based image restoration \cite{dong2011image,dong2011centralized,dong2013nonlocally}. Most recently, nonlocal sparsity has been connected with the powerful Gaussian scalar mixture (GSM) model \cite{portilla2003image} leading to the state-of-the-art performance in image restoration \cite{dong2015image}.

In learning-based approaches, deep neural network (DNN) techniques have attracted increasingly more attention and shown significant improvements in various low-level vision applications including superresolution (SR) and restoration \cite{dong2016image,kim2016accurate,kim2016deeply,wang2015deep,dong2016accelerating,shi2016real}. In \cite{cui2014deep}, stacked collaborative auto-encoders are used to gradually recover a high-resolution (HR) image layer by layer; in \cite{osendorfer2014image}, a SR method using predictive convolutional sparse coding and deconvolution network was developed. Multiple convolutional neural network \cite{dong2016image,kim2016accurate,kim2016deeply} have been proposed to directly learn the nonlinear mapping between low-resolution (LR) and high-resolution (HR) images; and multi-stage trainable nonlinear reaction diffusion network has also been proposed for image restoration \cite{chen2016trainable}. Moreover, most recent studies have shown that deeper neural network can lead to even better SR performance \cite{kim2016accurate,kim2016deeply}. However, it should be noted that the DNN approach \cite{dong2016image,kim2016accurate,kim2016deeply} still performs poorly on some particular sample images (e.g., if certain texture information is absent in the training data). Such mismatch between training and testing data is a fundamental limitation of all learning-based approaches.

One possible remedy for overcoming the above limitation is to explore somewhere between - i.e., a \emph{hybrid} approach combining the best of both worlds. Since training data and degraded image respectively contain supplementary (external and internal) prior information, it is natural to combine  them for image restoration. The key challenge is how to pursue such a hybrid approach in a principled manner. Inspired by the previous work connecting DNN with sparse coding (e.g., \cite{gregor2010learning} and \cite{wang2015deep}), we propose a Structured Analysis Sparse Coding (SASC) framework to jointly exploit the prior in both external and internal sources. Specifically, an external structured sparse prior is learned from training data via a deep convolutional neural network (in a similar way to previous learning-based approaches); meantime another internal structured sparse prior is estimated from the degraded image (similar to previous model-based approaches). Two structured sparse priors will be combined to produce a hybrid prior incorporating the knowledge from both domains. To manage the computational complexity, we have developed a novel framework of implementing hybrid structured sparse coding processes by deep convolutional neural networks. Experimental results have shown that the proposed hybrid image restoration method performs comparably with and often better than the current state-of-the-art techniques.

\section{Related Work}

\subsection{Sparse models for image restoration}

Generally speaking, sparse models can be classified into \emph{synthesis} models and \emph{analysis} models \cite{nam2013cosparse}. Synthesis sparse models assume that image patches can be represented as linear combinations of a few atoms from a dictionary. Let $\ty=\tH\tx+\tn$ denote the degraded image, where $\tH\in\mathbb{R}^{N\times M}$ is the observation matrix (e.g. blurring and down-sampling) and $\tn\in\mathbb{R}^N$ is the additive Gaussian noise. Then synthesis sparse model based image restoration can be formulated as Eq. (\ref{Syn_sparse})
\begin{small}
\begin{equation}
(\tx, \valpha_i)=\argmin_{\tx, \valpha_i} ||\ty-\tH\tx||_2^2 + \eta\sum_i \{||\tR_i\tx-\tD\valpha_i||_2^2 + \lambda||\valpha_i||_1 \},  \label{Syn_sparse}
\end{equation}
\end{small}
%\begin{align}(\tx, \valpha_i)=\argmin_{\tx, \valpha_i} ||\ty-\tH\tx||_2^2 + \nonumber\\
%&\eta\sum_i \{||\tR_i\tx-\tD\valpha_i||_2^2 + \lambda||\valpha_i||_1 \},  \label{Syn_sparse}
%\end{align}
where $\tR_i$ denote the matrix extracting patches of size $\sqrt{n}\times \sqrt{n}$ at position $i$ and $\tD\in\mathbb{R}^{n\times K}$ is the dictionary. The above optimization problem can be solved by alternatively optimizing $\valpha_i$ and $\tx$. The $\ell_1$ norm minimization problem in Eq. (\ref{Syn_sparse}) requires many iterations and is typically computational expensive.

Alternatively, analysis sparse model (ASC) \cite{nam2013cosparse} assumes that image patches are sparse in a transform domain- i.e., for a given dictionary $\tW\in\mathbb{R}^{K\times n}$ of analysis, $||\tW\tx_i||_0\ll K$ is sparse. With the ASC model, the unknown image can be recovered by solving
%\begin{equation}
%\begin{split}
%(\tx,\valpha_i) = \argmin_{\tx, \valpha_i} &||\ty-\tH\tx||_2^2 + \eta\sum_i \{||\tW(\tR_i\tx)-\valpha_i||_2^2 +\lambda||\valpha_i||_1 \}.  \label{Ana_sparse}
%\end{split}
%\end{equation}
\begin{align}(\tx,\valpha_i) = \argmin_{\tx, \valpha_i} &||\ty-\tH\tx||_2^2 + \nonumber\\
&\eta\sum_i \{||\tW(\tR_i\tx)-\valpha_i||_2^2 +\lambda||\valpha_i||_1 \}.  \label{Ana_sparse}
\end{align}

Note that if image patches are extracted with maximum overlapping along both horizontal and vertical directions, the transformation of each patches can be implemented by the convolution with the set of filters $\tw_k, k=1,2, \cdots, K$ with $\tx$- i.e.,
%\begin{equation}
%\begin{split}
%(\tx,\tz_k) = \argmin_{\tx, \tz_k} &||\ty-\tH\tx||_2^2 + \eta\sum_{k=1}^K \{||\tw_k*\tx-\tz_k||_2^2 +\lambda||\tz_k||_1 \}.  \label{Ana_sparse2}
%\end{split}
%\end{equation}
\begin{align}(\tx,\tz_k) = \argmin_{\tx, \tz_k} &||\ty-\tH\tx||_2^2 + \nonumber\\
&\eta\sum_{k=1}^K \{||\tw_k*\tx-\tz_k||_2^2 +\lambda|| \tz_k||_1 \}.  \label{Ana_sparse2}
\end{align}
$z_k$ represents sparse feature map corresponding to filter $w_k$. Compared with the synthesis sparse model, sparse codes or feature maps in Eq. (\ref{Ana_sparse}) and (\ref{Ana_sparse2}) can be solved in a closed-form solution, leading to significant reduction in computational complexity.

\subsection{Connecting sparsity with neural networks}

Recent studies have shown that sparse coding problem can be approximately solved by a neural network \cite{gregor2010learning}. In \cite{gregor2010learning}, a feed-forward neural network, which mimics the process of sparse coding, is proposed to approximate the sparse codes $\valpha_i$ with respect to a given synthesis dictionary $\tD$. By joint learning all model parameters from training dataset, good approximation of the underlying sparse codes can be obtained. In \cite{wang2015deep}, the connection between sparse coding and neural networks has been further extended for the application of image SR. Sparse coding (SC) based neural network is designed to emulate sparse coding based SR process - i.e., sparse codes of LR patches are first approximated by a neural network and then used to reconstruct HR patches with a HR synthesis dictionary. By jointly training all model parameters, SC-based neural network can achieve much better results than conventional SC-based methods. The fruitful connection between sparse coding and neural networks also inspires us to combine them in a more principled manner in this paper.

\section{Structured analysis sparse coding (SASC) for image restoration}

The analysis SC model of Eq. (\ref{Ana_sparse}) and (\ref{Ana_sparse2}) has the advantage of computational efficiency when compared to the synthesis SC model. However, $\ell_1$-norm based SC model ignores the correlation among sparse coefficients, leading to unsatisfactory results. Similar to previous works of nonlocal sparsity \cite{dong2011centralized,dong2013nonlocally}, a structured ASC model for image restoration can be formulated as
%\begin{equation}
%\begin{split}
%(\tx,\tz_k) = \argmin_{\tx, \tz_k} &||\ty-\tH\tx||_2^2 + \eta\sum_{k=1}^K \{||\tw_k*\tx-\tz_k||_2^2 + \lambda||\tz_k-\vmu_k||_1 \},  \label{Stru_ASC}
%\end{split}
%\end{equation}
\begin{align}(\tx,\tz_k) = \argmin_{\tx, \tz_k} &||\ty-\tH\tx||_2^2 + \nonumber\\
&\eta\sum_{k=1}^K \{||\tw_k*\tx-\tz_k||_2^2 + \lambda||\tz_k-\vmu_k||_1 \},  \label{Stru_ASC}
\end{align}

where $\vmu_k$ denotes the new nonlocal prior of the feature map (note that when $\vmu_k=\bm{0}$ the structured ASC model reduces to the conventional ASC model in Eq. (\ref{Ana_sparse2})). The introduction of $\vmu_k$ to sparse prior has the potential of leading to significant improvement of the estimation of sparse feature map $\tz_k$, which bridges the two competing approaches (model-based vs. learning-based).

The objective function of Eq. (\ref{Stru_ASC}) can be solved by alternatively optimizing $\tx$ and $\tz_k$. With fixed feature maps $\tz_k$, the restored image $\tx$ can be updated by computing
\begin{equation}
\tx = (\tH^{\top}\tH+\eta\sum_k\tW_k^{\top}\tW_k)^{-1}(\tH^{\top}\ty + \eta\sum_k\tW_k^{\top}\tz_k), \label{Cal_x}
\end{equation}
where $\tW_k\tx = \tw_k*\tx$ denotes the 2D convolution with filter $\tw_k$. Since the matrix to be inverted in Eq. (\ref{Cal_x}) is very large, it is impossible to compute Eq. (\ref{Cal_x}) directly. Instead, it can be computed by the iterative conjugated gradient (CG) algorithm, which requires many iterations. Here, instead of computing an exact solution of the $\tx$-subproblem, we propose to update $\tx$ with a single step of gradient descent of the objective function for an inexact solution, as
\begin{equation}
\begin{split}
\tx^{(t+1)} &= \tx^{(t)} - \delta[\tH^{\top}(\tH\tx^{(t)} - \ty) + \eta\sum_k\tW_k^{\top}(\tW_k\tx^{(t)}-\tz_k)] \\
&=\tA\tx^{(t)} + \delta\tH^{\top}\ty + \delta\eta\sum_k\tW_k^{\top}\tz_k, \label{Update_x}
\end{split}
\end{equation}
where $\tA=\tI-\delta\tH^{\top}\tH-\delta\eta\sum_k\tW_k^{\top}\tW_k$, $\delta$ is the predefined step size, and $\tx^{(t)}$ denotes the estimate of the whole image $\tx$ at the $t$-th iteration. As will be shown later, the update of $\tx^{(t)}$ can be efficiently implemented by convolutional operations. With fixed $\tx$, the feature maps can be updated via
\begin{equation}
\tz_k = \mathcal{S}_{\lambda/2}(\tw_k*\tx - \vmu_k) + \vmu_k, \label{Cal_features}
\end{equation}
where $\mathcal{S}_{\lambda}(\cdot)$ denotes the soft-thresholding operator with a threshold of $\lambda$. Now the question boils down tos how to accurately estimate $\vmu_k$. In the following subsections, we propose to learn the structured sparse prior from both training data (external) and the degraded image (internal).

\subsection{Prior learning from training dataset}

\begin{figure}
\centering
  \includegraphics[width=0.8\linewidth]{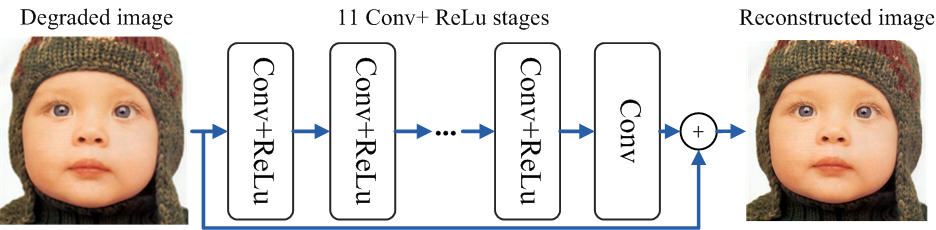}
\caption{The structure of the CNN for prior learning. The CNN contains 11 convolution layer with ReLu nonlinear activation function. For each convolution layer, 64 filters of size $3\times 3$ are used. The degraded image is fed into the network to get an initial estimate of the original image.}
\label{fig:cnn}
\end{figure}

For a given observation image $\ty$, we target at learning the feature maps $\tz_k$ of a desirable restored image $\tx$ with respect to filters $\tw_k$. Without the loss of generality, the learning function can be defined as follows
\begin{equation}
\hat{\tZ}   =  G(\ty; \tTheta),
\end{equation}
where $\hat{\tZ}=[\hat{\tz}_1,\hat{\tz}_2,\cdots,\hat{\tz}_K]$ and $G(\cdot)$ denotes the learning function parameterized by $\tTheta$. Considering the strong representing abilities of convolutional neural networks (CNN), we choose to learn $\tz_k$ on a deep CNN (DCNN). We have found that directly learning a set of feature maps $\tz_k$ with respect to $\tw_k$ is unstable; instead, we propose to first learn the desirable restored image $\hat{\tx}$ and then compute the feature maps via $\hat{\tz}_k=\tw_k*\hat{\tx}$. Generally speaking, any existing DCNN can be used for an initial estimate of $\tx$. The architecture of DCNN (as shown in Fig. \ref{fig:cnn}) is similar to that of \cite{dong2016image}. However, different from \cite{dong2016image}, convolution filters of smaller size and more convolution layers are used for better estimation performance. The CNN contains $12$ convolution layer, each of which uses 64 filters sized by $3\times 3\times 64$. The last layer uses a single filter of size $3\times 3$ for reconstruction. A shortcut or skip connection (not shown in the figure) exists from input to output implementing the concept of deep residue learning (similar to \cite{kim2016deeply}). The objective function of DCNN training can be formulated as
\begin{equation}
\tTheta = \argmin_{\tTheta}\sum_i ||CNN(\ty_i;\tTheta) - \tx_i||_2^2
\end{equation}
where $\ty_i$ and $\tx_i$ denotes the observed and target training image pairs and $CNN(\ty_i;\tTheta)$ denotes the output of CNN with parameters $\tTheta$. All network parameters are optimized through the back-propagation algorithm. After the estimation of $\tx$, the set of feature maps can be estimated by convoluting it with a set of analysis filters $\tw_k$- i.e., $\hat{\tz_k} = \tw_k*\hat{\tx}, k=1,2,\cdots,K$.

\subsection{Prior learning by exploiting nonlocal self-similarity}

In addition to externally learning the prior feature maps via CNN, we can also obtain the estimates of $\tz_k$ from an \emph{internal} estimate of the target image.  Let $\hat{\tx}_i$ denote the patch of size $\sqrt{n}\times\sqrt{n}$ extracted at position $i$ from an initial estimate $\hat{\tx}$; then sparse codes of $\hat{\tx}_i$ can be computed as $\hat{\tz}_{i,k} = \tw_k^{\top}\hat{\tx}_i$. Considering that the natural images contain rich self-repetitive patterns, a better estimate of $\tz_{i,k}$ can be obtained by a weighted average of the sparse codes over similar patches. Let $\hat{\tx}_{i_l}, l=1,2,\cdots,L$ denote the set of similar patches that are within the first $L$-th closest matches and $G_i={i_1,i_2\cdots,i_L}$ denote the collection of the positions corresponding to those similar patches. A nonlocal estimate of $\hat{\tz}_{i,k}$ can be calculated as
\begin{equation}
\tilde{\tz}_{i,k} = \sum_{l=1}^L w_{i_l}\tw_k^{\top}\hat{\tx}_{i_l} = \tw_k^{\top}\tilde{\tx}_{i}, \label{NL_feature}
\end{equation}
where $w_{i_l}=\frac{1}{c}\exp(-||\hat{\tx}_{i_l}-\hat{\tx}_{i}||/h)$, $c$ is the normalization constant, $h$ is the predefined constant, and $\tilde{\tx}_i=\sum_{l=1}^Lw_{i_l}\hat{\tx}_{i_l}$. From Eq. (\ref{NL_feature}), we can see that a nonlocal estimate of the sparse codes can be obtained by first computing the nonlocal estimate of the target image followed by a 2D convolution with the filters $\tw_k$.

By combining the estimate obtained by CNN and nonlocal estimation, an improved hybrid prior of the feature maps can be obtained by
\begin{equation}
\vmu_k = \delta \hat{\tz}_{k} + (1-\delta)\tilde{\tz}_k, \label{Cal_mu}
\end{equation}
where $0<\delta<1$ is a preselected constant. The overall structured analysis sparse coding (SASC) with prior learning for image restoration is summarized in \textbf{Algorithm 1}. We note that \textbf{Algorithm 1} usually requires dozens of iterations for converging to a satisfactory result. Hence, the computational cost of the proposed SASC model is high; meanwhile, the analysis filters $\tw_k$ used in \textbf{Algorithm 1} are kept fixed.
%which has to be learned in an off-line learning stage or selected as an off-the-shelf analysis dictionary (e.g., DCT or wavelet transforms).
A more computationally efficient implementation is to approximate the proposed SASC model by a deep neural network. Through end-to-end training, we can jointly optimize the parameters $\eta$, $\lambda$ and the analysis filters $\tw_k$ as will be elaborated next.

\setenumerate[1]{itemsep=0pt,partopsep=0pt,parsep=\parskip,topsep=5pt}
\setitemize[1]{itemsep=0pt,partopsep=0pt,parsep=\parskip,topsep=5pt}
\setdescription{itemsep=0pt,partopsep=0pt,parsep=\parskip,topsep=5pt}
\begin{algorithm}[t]
\textbf{Initialization}:\
\begin{itemize}
\item[(a)] Set parameters $\eta$ and $\lambda$;
\item[(b)] Compute the initial estimate $\hat{\tx}^{(0)}$ by the CNN;
\item[(c)] Group a set of similar patches $G_i$ for each patch $\hat{\tx}_{i}$ using $\hat{\tx}^{(0)}$;
\item[(c)] Compute the prior feature maps $\vmu_k$ using Eq. (\ref{Cal_mu});
\end{itemize}
\textbf{Outer loop}: Iteration over $t=1,2,\cdots,T$
\begin{itemize}
\item[(a)] Compute the feature maps $\tz_k^{(t)}, k=1,\cdots,K$ using Eq. (\ref{Cal_features});
\item[(b)] Update the HR image $\hat{\tx}^{(t)}$ via Eq. (\ref{Update_x});
\item[(c)] Update $\vmu_k$ via Eq. (\ref{Cal_mu}) based on $\hat{\tx}^{(t)}$;
\end{itemize}
\textbf{Output}: $\tx^{(t)}$.
\caption{Image SR with structured ASC}
\label{alg:Alg1}
\end{algorithm}

\section{Network implementation of SASC for image restoration}

\begin{figure*}
\centering
  \includegraphics[width=1\linewidth]{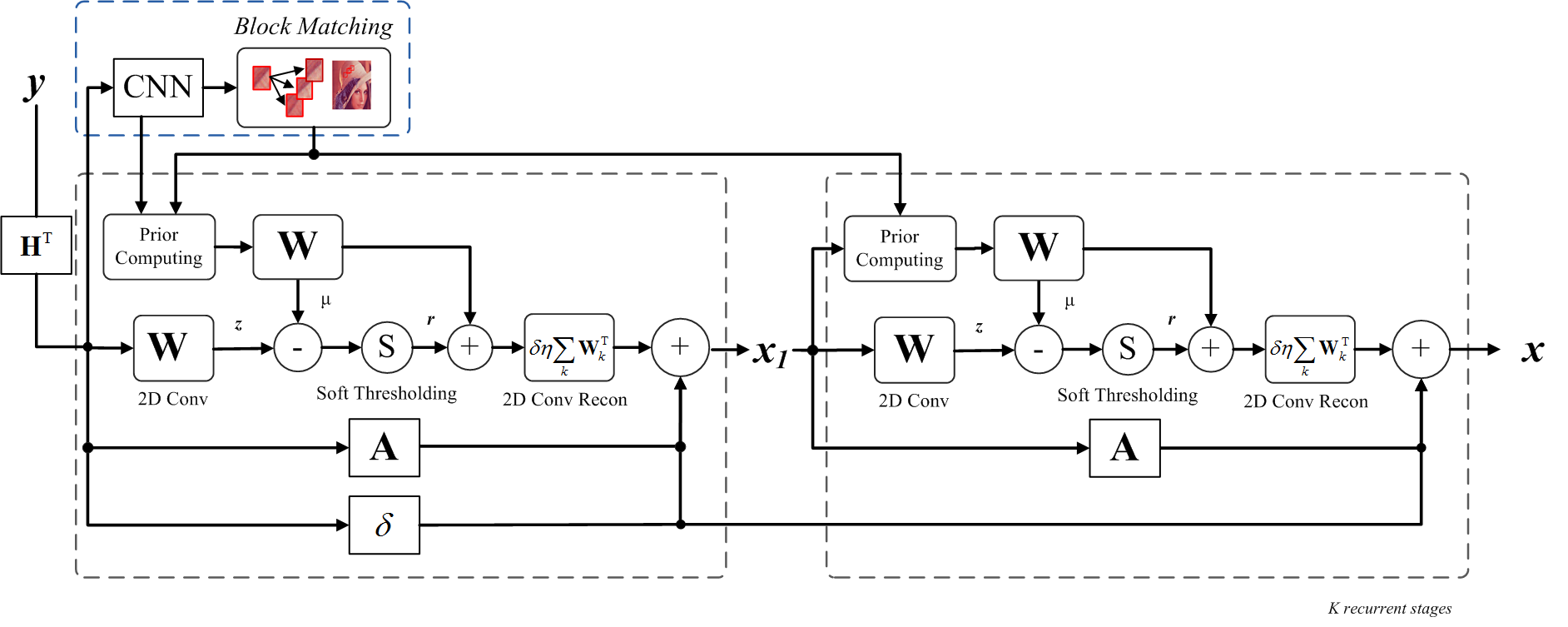}
\caption{The structure of the proposed SASC network for image restoration. The whole architecture consists of CNN sub-network and SASC sub-network. Degraded image or intermediate result combine with CNN estimates, feed into multiple SASC recurrent stages to get the final reconstructed image.}
\label{Fig:NN_structure}
\end{figure*}

The main architecture for network implementation of SASC is shown in Fig. (\ref{Fig:NN_structure}), which mimics the iterative steps of \textbf{Algorithm 1}.
%As shown in Fig. (\ref{Fig:NN_structure}), the network consists of two sub-networks, i.e., the prior learning sub-network and the SASC sub-network.
As shown in Fig. (\ref{Fig:NN_structure}), the degraded observation image $\ty$ goes through the CNN for an initial estimate of the target image, which will then be used for grouping similar patches and computing prior feature maps. Let $G_i$ denote the set of similar patch positions for each exemplar patch $\hat{\tx}_i$ (for computational simplicity, $G_i$ will not be updated during the iterative processing).

The initial estimate obtained via CNN and the set of similar patch positions $G_i$ are then fed into the SASC network that contains $k$ recurrent stages to reconstruct the target image. The SASC network exactly mimics the process of alternatively updating of the feature maps $\tz_k$ and the HR image $\tx$ as shown in Eq. (\ref{Cal_features}) and (\ref{Update_x}). The degraded image $\ty$ (after bicubic interpolation if down-sampling is involved) first goes through a convolution layer for sparse feature maps $\tz_k$, which will then be predicted by the learned prior feature maps $\vmu_k$. The residuals of the predicted feature maps, denoted by $\tr_k$, will go through a nonlinear soft-thresholding layer. Similar to \cite{wang2015deep}, we can write the soft-thresholding operator as
\begin{equation}
s_{\tau_i}(r_i) = \text{Sign}(r_i)r_i(|r_i|/\tau_i-1)_+,
\end{equation}
where $\tau_i$ denotes a tunable threshold. Note that the soft-thresholding layer can be implemented as two linear layers and a unit-threshold layer. After soft-thresholding layer, the learned prior feature maps $\vmu_k$ are added back to the output of soft-thresholding layer. The updated feature maps $\tz_k$ then go through a reconstruction layer with a set of 2D convolution filters- i.e., $\sum_k\tW_k^{\top}\tz_k$. The final output of the reconstruction layer is further added with the preprocessed degraded image- i.e., denoted as $\tH^{\top}\ty$. Finally, the weighted intermediate result of reconstructed HR image is fed into a linear layer parameterized by matrix $\tA$. Note that $\tA$ corresponds to the matrix - i.e.,
\begin{equation}
\tA=\tI-\delta\tH^{\top}\tH-\delta\eta\sum_k\tW_k^{\top}\tW_k.
\end{equation}
Note that $\sum_k(\tW_k^{\top}\tW_k\tx)$ can be efficiently computed by first convoluting $\tx$ with 2D filters and adding up the resulting feature maps - i.e., $\sum_k(\tw_k^{\top}*(\tw_k*\tx))$. For typical degradation matrices $\tH$, $\bar{\tH}=\tH^{\top}\tH$ can also be efficiently computed by convolutional operations. For image denoising, $\bar{\tH}=\tI$. For image deblurring, the matrix-vector multiplication $\tH^{\top}(\tH\tx)$ can be simply implemented by two convolutional operations. For image super-resolution, we consider two typical downsampling operators, i.e., the Gaussian downsampling and the bicubic downsampling. For Gaussian downsampling, \tH=\tD\tB, where $\tD$ and $\tB$ denote the downsampling and Gaussian blur matrices, respectively. In this case, $\ty=\tH\tx$ can be efficiently computed by first convoluting $\tx$ with the corresponding Gaussian filter followed by subsampling, whereas $\tH^{\top}\ty$ can also be efficiently computed by first upsampling $\ty$ with zero-padding followed by convolution with the transposed Gaussian filter. For bicubic downsampling, we simply use the bicubic interpolator function with scaling factor $s$ and $1/s$ ($s=2,3,4$) to implement $\tH^{\top}\ty$ and $\tH\tx$, respectively. Note that all convolutional filters and the scale variables involved in the linear layer $\tA$ can be discriminately learned through end-to-end training. After going through the linear $\tA$, we obtain the reconstructed image; for better performance, such SASC sub-network can be repeated $K$ times.

\begin{figure}
\centering
  \includegraphics[width=1\linewidth]{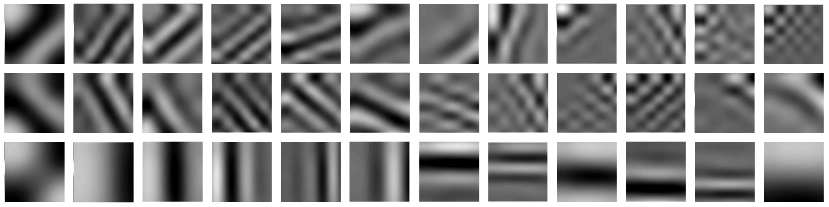}
\caption{Visualization of some of the learned analysis filters in first SASC stage. It can be infer from this figure that the filters has different responses to the edge features of different directions and frequencies.}
\label{Fig:filters}
\end{figure}

In summary, there are totally $5$ trainable layers in each stage of our proposed network: two convolution layers $\tW$, one reconstruction layer parameterized with $\sum_k\tW_k^{\top}\tz_k$, one nonlinear soft-thresholding layer, and one linear layer $\tA$. Parameters at different stages are not same; but the $i$-th stage of diffenent networks share the same weights $W^{(i)}$. Mean square error is used as the cost function to train the network, and the overall objective function is given by
\begin{equation}
\tTheta = \argmin_{\tTheta}\sum_i ||SASC(\ty_i;\tTheta) - \tx_i||_2^2
\end{equation}
where $\tTheta$ denotes the set of parameters and $SASC(\ty_i;\tTheta)$ denotes the reconstructed image by the network with parameters $\tTheta$. To train the network, the ADAM optimizer \cite{} with setting $\beta_1=0.9$ and $\beta_2=0.999$ and $\epsilon=10^{-8}$ is used.
Note that to facilitate training, we separately train the CNN network and the SASC network. Some examples of the learned convolution filters are shown in Fig. (\ref{Fig:filters}).

\section{Experimental results}

To verify the performance of the proposed method, several image restoration experiments have been conducted, including denoising, deblurring and super-resolution. In all experiments, we empirically set $K=5$ stages for the proposed SASC network. To gain deeper insight toward the proposed SASC network, we have implemented several variants of the proposed SASC network. The first variant is the analysis sparse coding (ASC) network without CNN and self-similarity prior learning. The second variant of the proposed method is the SASC network with self-similarity prior, which estimate $\vmu_k$ from intermediately recovered HR image (without using CNN sub-network), which is denoted as SASC-SS method. We also present the image restoration results of the CNN sub-network, which consists of 12 convolutional layers with ReLU nonlinearity and $3\times 3\times 64$ kernels. The proposed SASC network with CNN and self-similarity prior learning is denoted as SASC-CNN-SS method. To train the networks, we have adopted three training sets: the train400 dataset used in \cite{Zhang:TIP17} for image denoising/deblurring, the 91 training images used in \cite{yang2010image} and the BSD200 dataset for image super-resolution.

\subsection{Image denoising}

\begin{figure*}[!tbh]
\renewcommand{\arraystretch}{0.4}
\centering
\subfigure[]{
    \includegraphics[width=0.07\textwidth]{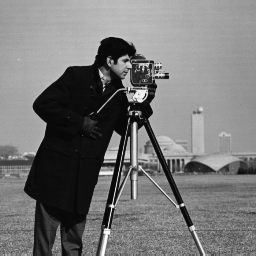}
    }\hspace{-0.6em}
\subfigure[]{
    \includegraphics[width=0.07\textwidth]{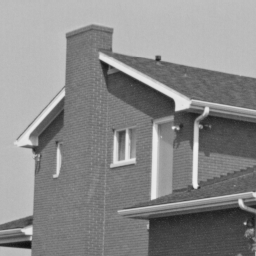}
    }\hspace{-0.6em}
\subfigure[]{
    \includegraphics[width=0.07\textwidth]{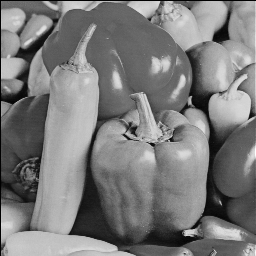}
    }\hspace{-0.6em}
\subfigure[]{
    \includegraphics[width=0.07\textwidth]{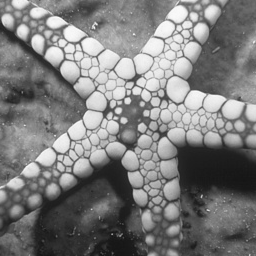}
    }\hspace{-0.6em}
\subfigure[]{
    \includegraphics[width=0.07\textwidth]{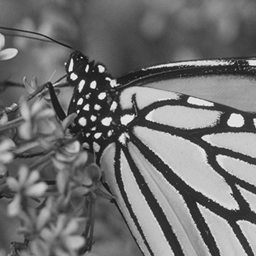}
    }\hspace{-0.6em}
\subfigure[]{
    \includegraphics[width=0.07\textwidth]{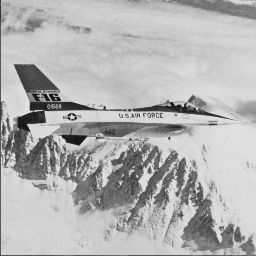}
    }\hspace{-0.6em}
\subfigure[]{
    \includegraphics[width=0.07\textwidth]{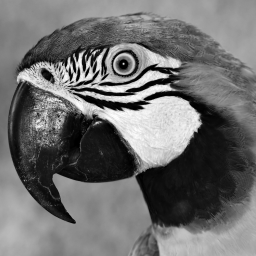}
    }\hspace{-0.6em}
\subfigure[]{
    \includegraphics[width=0.07\textwidth]{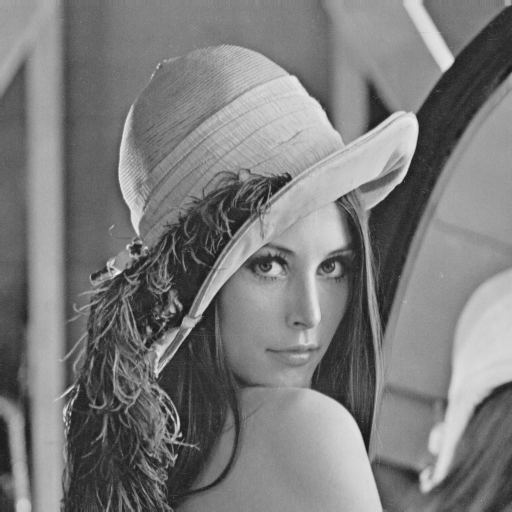}
    }\hspace{-0.6em}
\subfigure[]{
    \includegraphics[width=0.07\textwidth]{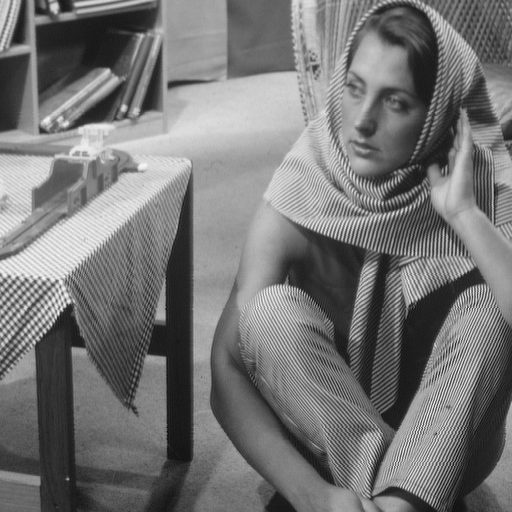}
    }\hspace{-0.6em}
\subfigure[]{
    \includegraphics[width=0.07\textwidth]{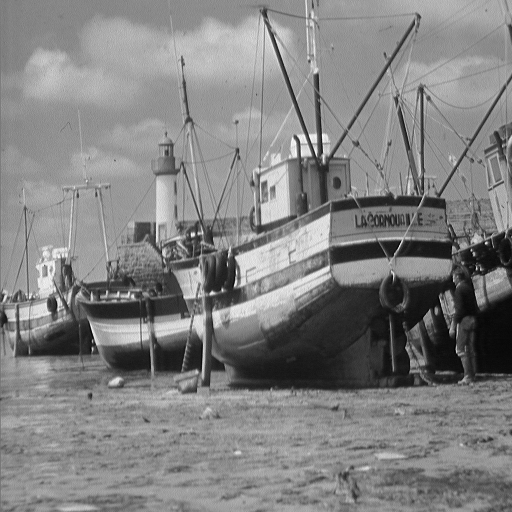}
    }\hspace{-0.6em}
\subfigure[]{
    \includegraphics[width=0.07\textwidth]{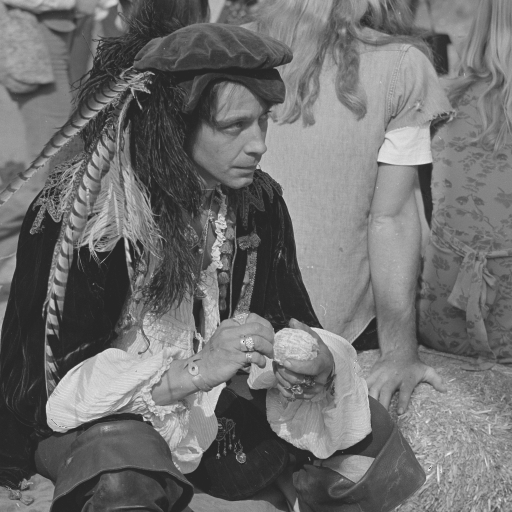}
    }\hspace{-0.6em}
\subfigure[]{
    \includegraphics[width=0.07\textwidth]{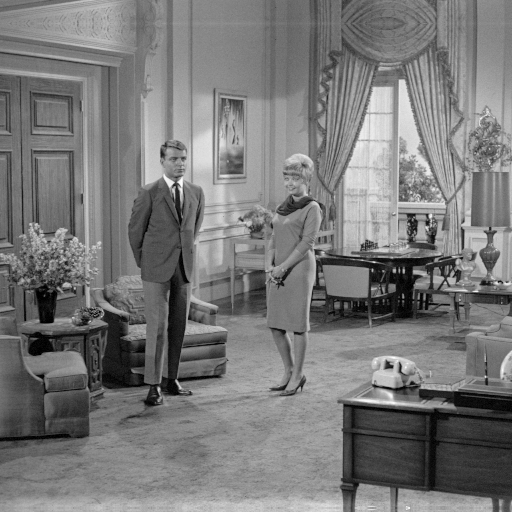}
    }
    \\      \caption{The test images used for image denoising/deblurring. From left to right: \textit{C.Man}, \textit{House}, \textit{Peppers}, \textit{Starfish}, \textit{Monarch}, \textit{Airplane}, \textit{Parrot}, \textit{Lena}, \textit{Barbara}, \textit{Boat}, \textit{Man}, and \textit{Couple}. }
\label{fig:set12}
\end{figure*}

In our experiment, we have extracted patches of size $40\times 40$ from the train400 dataset \cite{Zhang:TIP17} and used argumentation with flip and rotations to generate $6000\times 128$ patches as the training data. The commonly used $12$ images used in \cite{BM3D} (as shown in Fig. \ref{fig:set12}) were used as the test set. The BSD68 dataset was also used as a benchmark dataset. The average PSNR and SSIM results of the variants of the proposed SASC methods on the two sets are shown in Table \ref{Variants-den}. From Table \ref{Variants-den}, one can see that by incorporating the nonlocal self-similarity prior, the SASC-SS method outperforms the ASC method; by integrating both CNN (external) and nonlocal self-similarity (internal) priors, the proposed SASC-CNN-SS method further improves the denoising performance. \emph{Similar observations have also been made for image deblurring and super-resolution}. Due to the limited page spaces, here we only show the comparison studies of the variants of the proposed method for image denoising.

\begin{figure*}[!tbh]
\renewcommand{\arraystretch}{0.4}
\centering
\subfigure[]{
    \includegraphics[width=0.28\textwidth]{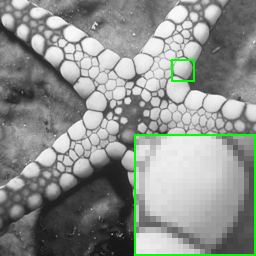}
    }\hspace{-0.8em}
\subfigure[]{
    \includegraphics[width=0.28\textwidth]{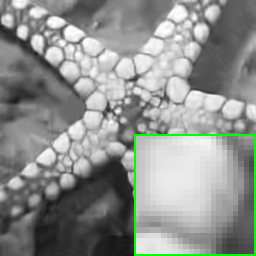}
    }\hspace{-0.8em}
\subfigure[]{
    \includegraphics[width=0.28\textwidth]{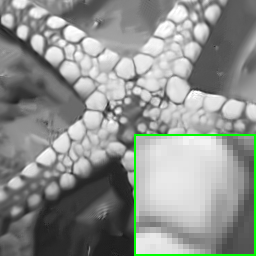}
    }\hspace{-0.8em}
\subfigure[]{
    \includegraphics[width=0.28\textwidth]{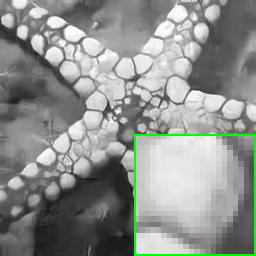}
    }\hspace{-0.8em}
\subfigure[]{
    \includegraphics[width=0.28\textwidth]{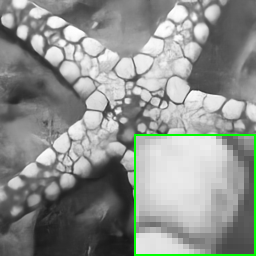}
    }\hspace{-0.8em}
\subfigure[]{
    \includegraphics[width=0.28\textwidth]{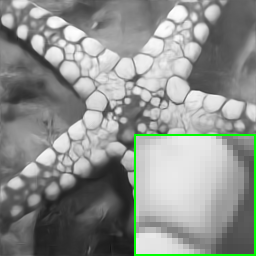}
    }\hspace{-0.8em}

     \caption{Denoising results of noise level of 50. (a) Parts of the original images of ``starfish'' in Set12; (b) BM3D\cite{BM3D}(PSNR=25.04dB); (c) WNNM\cite{WNNM}(PSNR=25.43dB); (d) TNRD\cite{TNRD}(PSNR=25.42dB);(e) DnCNN\cite{Zhang:TIP17}(PSNR=25.70dB); (f) \bf{Proposed method(PSNR=26.46dB)}}
\label{den-Barbara}
\end{figure*}

\begin{figure*}[!tbh]
\renewcommand{\arraystretch}{0.4}
\centering

\subfigure[]{
    \includegraphics[width=0.18\textwidth]{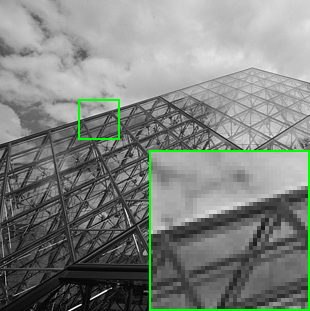}
    }\hspace{-0.8em}
\subfigure[]{
    \includegraphics[width=0.18\textwidth]{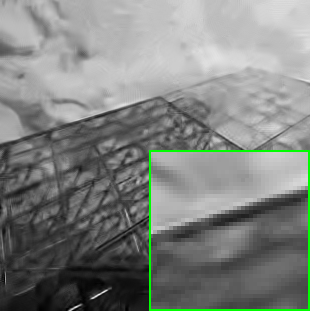}
    }\hspace{-0.8em}
\subfigure[]{
    \includegraphics[width=0.18\textwidth]{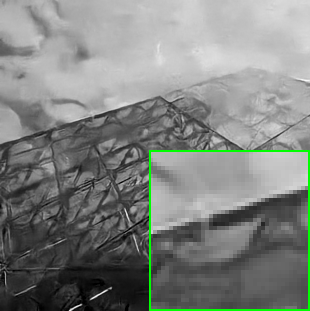}
    }\hspace{-0.8em}
\subfigure[]{
    \includegraphics[width=0.18\textwidth]{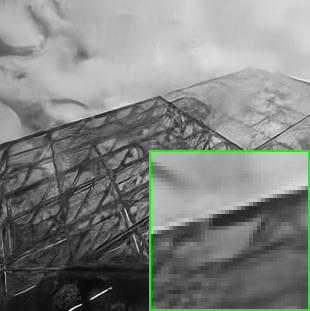}
    }\hspace{-0.8em}
\subfigure[]{
    \includegraphics[width=0.18\textwidth]{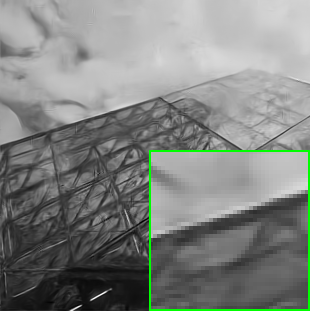}
    }\hspace{-0.8em}

    \caption{Denoising results of noise level of 50. (a) Parts of the original images of ``test044'' in BSD68; (b) BM3D\cite{BM3D}(PSNR=23.65dB); (c) TNRD\cite{TNRD}(PSNR=24.05dB);(d) DnCNN\cite{Zhang:TIP17}(PSNR=24.35dB); (e)\textbf{Proposed method(PSNR=24.89dB)}}
\label{den-test044}
\end{figure*}

We have also compared the proposed method with several popular denoising methods including model-based denoising methods (BM3D\cite{BM3D}, EPLL\cite{Zoran:ICCV11}, and WNNM \cite{WNNM}) and two deep learning based methods (TNRD \cite{TNRD} and DnCNN-S\cite{Zhang:TIP17}). Table \ref{den-set12} shows the PSNR results of the competing methods on 12 test images. It can be seen that the proposed method performs much better than other competing methods. Specifically, the proposed method outperforms current state-of-the-art DnCNN-S \cite{Zhang:TIP17} by up to $0.56dB$ on the average.  Parts of the denoised images by different methods are shown in Figs. \ref{den-Barbara}-\ref{den-test044}. It can be seen that the proposed method produces better visually pleasant results, as can be clearly observed in the regions of self-repeating patterns (edges and textures). %The PSNR results on the BSD68 dataset are shown in Table \ref{den-BSD68}, from which one can see that the proposed method still outperform other methods by a large margin.

\begin{table*}[tbh]
\centering
\caption{Average PSNR and SSIM results of the variants of the proposed denoising method}
\label{Variants-den}
%\begin{tabular}{|c|c|c|c|c|c|c|}
%\hline
\begin{tabular}{!{\vrule width1.2pt}c!{\vrule width1.2pt}c!{\vrule width1.2pt}c
|c|c|c|c|c|c|c|c|c!{\vrule width1.2pt}}
\Xhline{1.2pt}

\multirow{2}{*}{} & \multicolumn{3}{c|}{Set12}                                                                                                                                               & \multicolumn{3}{c|}{BSD68}                                                                                                                                               \\ \cline{2-7}
                  & $\sigma=15$                                                     & $\sigma=25$                                                     & $\sigma=50$                                                     & $\sigma=15$                                                     & $\sigma=25$                                                     & $\sigma=50$                                                     \\ \hline
ASC               & \begin{tabular}[c]{@{}c@{}}32.60\\0.8928\end{tabular} & \begin{tabular}[c]{@{}c@{}}30.30\\0.8470\end{tabular} & \begin{tabular}[c]{@{}c@{}}27.01\\0.7400\end{tabular} & \begin{tabular}[c]{@{}c@{}}31.65\\0.8825\end{tabular} & \begin{tabular}[c]{@{}c@{}}29.11\\0.8097\end{tabular} & \begin{tabular}[c]{@{}c@{}}26.01\\0.6704\end{tabular} \\ \hline

SASC-SS               & \begin{tabular}[c]{@{}c@{}}32.98\\0.9016\end{tabular} & \begin{tabular}[c]{@{}c@{}}30.57\\0.8601\end{tabular} & \begin{tabular}[c]{@{}c@{}}27.35\\0.7669\end{tabular} & \begin{tabular}[c]{@{}c@{}}31.88\\0.8888\end{tabular} & \begin{tabular}[c]{@{}c@{}}29.36\\0.8243\end{tabular} & \begin{tabular}[c]{@{}c@{}}26.34\\0.7006\end{tabular} \\ \hline

CNN-Prior               & \begin{tabular}[c]{@{}c@{}}32.85\\0.8897\end{tabular} & \begin{tabular}[c]{@{}c@{}}30.38\\0.8394\end{tabular} & \begin{tabular}[c]{@{}c@{}}27.24\\0.7611\end{tabular} & \begin{tabular}[c]{@{}c@{}}31.75\\0.8839\end{tabular} & \begin{tabular}[c]{@{}c@{}}29.17\\0.8115\end{tabular} & \begin{tabular}[c]{@{}c@{}}26.23\\0.6924\end{tabular} \\ \hline

SASC-CNN-SS               & \begin{tabular}[c]{@{}c@{}}33.32\\0.9039\end{tabular} & \begin{tabular}[c]{@{}c@{}}30.99\\0.8673\end{tabular} & \begin{tabular}[c]{@{}c@{}}27.69\\0.7915\end{tabular} & \begin{tabular}[c]{@{}c@{}}32.03\\0.8870\end{tabular} & \begin{tabular}[c]{@{}c@{}}29.63\\0.8289\end{tabular} & \begin{tabular}[c]{@{}c@{}}26.66\\0.7254\end{tabular} \\ \hline

\end{tabular}
\end{table*}

\begin{table*}[tbh]
\centering
\caption{Results of proposed denoising method in Set12}
\setlength{\tabcolsep}{4.5pt}
\begin{tabular}{!{\vrule width1pt}l!{\vrule width1pt}l
!{\vrule width1pt}l!{\vrule width1pt}l!{\vrule width1pt}l
!{\vrule width1pt}l!{\vrule width1pt}l!{\vrule width1pt}l
!{\vrule width1pt}l!{\vrule width1pt}l!{\vrule width1pt}l
!{\vrule width1pt}l!{\vrule width1pt}l!{\vrule width1pt}l!{\vrule width1pt}}\Xhline{1.2pt}

\multirow{1}{*}{IMAGE}       & \multirow{1}{*}{C.Man} 	& \multirow{1}{*}{House} 	
& \multirow{1}{*}{Peppers} 		& \multirow{1}{*}{Starfish} 	& \multirow{1}{*}{Monar} 		& \multirow{1}{*}{Airpl} 	& \multirow{1}{*}{Parrot} 		 & \multirow{1}{*}{Lena} 		& \multirow{1}{*}{Barbara} 		& \multirow{1}{*}{Boat} 		& \multirow{1}{*}{Man} 			& \multirow{1}{*}{Couple} 	& \multirow{1}{*}{\textbf{Avg}} \\ \Xhline{1pt}

\multirowthead{1}{Noise Lv} & \multicolumn{13}{c!{\vrule width1pt}} {$\sigma=15$}
\\ \Xhline{1pt}
\multirowthead{1}{\cite{BM3D}}        & \multicolumn{1}{l|}{31.92}      & \multicolumn{1}{l|}{34.94}      & \multicolumn{1}{l|}{32.70}        & \multicolumn{1}{l|}{31.15}         & \multicolumn{1}{l|}{31.86}      & \multicolumn{1}{l|}{31.08}      & \multicolumn{1}{l|}{31.38}       & \multicolumn{1}{l|}{34.27}     & \multicolumn{1}{l|}{33.11}        & \multicolumn{1}{l|}{32.14}     & \multicolumn{1}{l|}{31.93}    & \multicolumn{1}{l!{\vrule width1pt}}{32.11}       & \multicolumn{1}{l!{\vrule width1pt}}{32.38}                 \\ \hline
\multirowthead{1}{\cite{WNNM}}        & \multicolumn{1}{l|}{32.18}      & \multicolumn{1}{l|}{35.15}      & \multicolumn{1}{l|}{32.97}        & \multicolumn{1}{l|}{31.83}         & \multicolumn{1}{l|}{32.72}      & \multicolumn{1}{l|}{31.40}      & \multicolumn{1}{l|}{31.61}       & \multicolumn{1}{l|}{34.38}     & \multicolumn{1}{l|}{33.61}        & \multicolumn{1}{l|}{32.28}     & \multicolumn{1}{l|}{32.12}    & \multicolumn{1}{l!{\vrule width1pt}}{32.18}       & \multicolumn{1}{l!{\vrule width1pt}}{32.70}                 \\ \hline
\multirowthead{1}{\cite{Zoran:ICCV11}}        & \multicolumn{1}{l|}{31.82}      & \multicolumn{1}{l|}{34.14}      & \multicolumn{1}{l|}{32.58}        & \multicolumn{1}{l|}{31.08}         & \multicolumn{1}{l|}{32.03}      & \multicolumn{1}{l|}{31.16}      & \multicolumn{1}{l|}{31.40}       & \multicolumn{1}{l|}{33.87}     & \multicolumn{1}{l|}{31.34}        & \multicolumn{1}{l|}{31.91}     & \multicolumn{1}{l|}{31.97}    & \multicolumn{1}{l!{\vrule width1pt}}{31.90}       & \multicolumn{1}{l!{\vrule width1pt}}{32.10}                 \\ \hline
\multirowthead{1}{\cite{TNRD}}        & \multicolumn{1}{l|}{32.19}      & \multicolumn{1}{l|}{34.55}      & \multicolumn{1}{l|}{33.03}        & \multicolumn{1}{l|}{31.76}         & \multicolumn{1}{l|}{32.57}      & \multicolumn{1}{l|}{31.47}      & \multicolumn{1}{l|}{31.63}       & \multicolumn{1}{l|}{34.25}     & \multicolumn{1}{l|}{32.14}        & \multicolumn{1}{l|}{32.15}     & \multicolumn{1}{l|}{32.24}    & \multicolumn{1}{l!{\vrule width1pt}}{32.11}       & \multicolumn{1}{l!{\vrule width1pt}}{32.51}                 \\ \hline
\multirowthead{1}{\cite{Zhang:TIP17}}       & \multicolumn{1}{l|}{\textbf{32.62}}      & \multicolumn{1}{l|}{35.00}      & \multicolumn{1}{l|}{33.29}        & \multicolumn{1}{l|}{32.23}         & \multicolumn{1}{l|}{33.10}      & \multicolumn{1}{l|}{31.70}      & \multicolumn{1}{l|}{31.84}       & \multicolumn{1}{l|}{34.63}     & \multicolumn{1}{l|}{32.65}        & \multicolumn{1}{l|}{32.42}     & \multicolumn{1}{l|}{32.47}    & \multicolumn{1}{l!{\vrule width1pt}}{32.47}       & \multicolumn{1}{l!{\vrule width1pt}}{32.87}                 \\ \hline

\multirowthead{1}{\textbf{Ours}}        & \multicolumn{1}{l|}{32.16}      & \multicolumn{1}{l|}{\textbf{35.51}}      & \multicolumn{1}{l|}{\textbf{33.87}}        & \multicolumn{1}{l|}{\textbf{32.67}}         & \multicolumn{1}{l|}{\textbf{33.30}}      & \multicolumn{1}{l|}{\textbf{31.98}}      & \multicolumn{1}{l|}{\textbf{32.21}}       & \multicolumn{1}{l|}{\textbf{35.19}}     & \multicolumn{1}{l|}{\textbf{33.92}}        & \multicolumn{1}{l|}{\textbf{32.99}}     & \multicolumn{1}{l|}{\textbf{32.93}}    & \multicolumn{1}{l!{\vrule width1pt}}{\textbf{33.08}}       & \multicolumn{1}{l!{\vrule width1pt}}{\textbf{33.31}}                 \\ \Xhline{1pt}

\multirowthead{1}{Noise Lv} & \multicolumn{13}{c!{\vrule width1pt}}{$\sigma=25$}                                                                                                                                                                                                                                                                                                                                                             \\ \Xhline{1pt}
\multirowthead{1}{\cite{BM3D}}        & \multicolumn{1}{l|}{29.45}      & \multicolumn{1}{l|}{32.86}      & \multicolumn{1}{l|}{30.16}        & \multicolumn{1}{l|}{28.56}         & \multicolumn{1}{l|}{29.25}      & \multicolumn{1}{l|}{28.43}      & \multicolumn{1}{l|}{28.93}       & \multicolumn{1}{l|}{32.08}     & \multicolumn{1}{l|}{30.72}        & \multicolumn{1}{l|}{29.91}     & \multicolumn{1}{l|}{29.62}    & \multicolumn{1}{l!{\vrule width1pt}}{29.72}       & \multicolumn{1}{l!{\vrule width1pt}}{29.98}                 \\ \hline
\multirowthead{1}{\cite{WNNM}}        & \multicolumn{1}{l|}{29.64}      & \multicolumn{1}{l|}{33.23}      & \multicolumn{1}{l|}{30.40}        & \multicolumn{1}{l|}{29.03}         & \multicolumn{1}{l|}{29.85}      & \multicolumn{1}{l|}{28.69}      & \multicolumn{1}{l|}{29.12}       & \multicolumn{1}{l|}{32.24}     & \multicolumn{1}{l|}{31.24}        & \multicolumn{1}{l|}{30.03}     & \multicolumn{1}{l|}{29.77}    & \multicolumn{1}{l!{\vrule width1pt}}{29.82}       & \multicolumn{1}{l!{\vrule width1pt}}{30.26}                 \\ \hline
\multirowthead{1}{\cite{Zoran:ICCV11}}        & \multicolumn{1}{l|}{29.24}      & \multicolumn{1}{l|}{32.04}      & \multicolumn{1}{l|}{30.07}        & \multicolumn{1}{l|}{28.43}         & \multicolumn{1}{l|}{29.30}      & \multicolumn{1}{l|}{28.56}      & \multicolumn{1}{l|}{28.91}       & \multicolumn{1}{l|}{31.62}     & \multicolumn{1}{l|}{28.55}        & \multicolumn{1}{l|}{29.69}     & \multicolumn{1}{l|}{29.63}    & \multicolumn{1}{l!{\vrule width1pt}}{29.48}       & \multicolumn{1}{l!{\vrule width1pt}}{29.63}                 \\ \hline
\multirowthead{1}{\cite{TNRD}}        & \multicolumn{1}{l|}{29.71}      & \multicolumn{1}{l|}{32.54}      & \multicolumn{1}{l|}{30.55}        & \multicolumn{1}{l|}{29.02}         & \multicolumn{1}{l|}{29.86}      & \multicolumn{1}{l|}{28.89}      & \multicolumn{1}{l|}{29.18}       & \multicolumn{1}{l|}{32.00}     & \multicolumn{1}{l|}{29.41}        & \multicolumn{1}{l|}{29.92}     & \multicolumn{1}{l|}{29.88}    & \multicolumn{1}{l!{\vrule width1pt}}{29.71}       & \multicolumn{1}{l!{\vrule width1pt}}{30.06}                 \\ \hline
\multirowthead{1}{\cite{Zhang:TIP17}}       & \multicolumn{1}{l|}{\textbf{30.19}}      & \multicolumn{1}{l|}{33.09}      & \multicolumn{1}{l|}{30.85}        & \multicolumn{1}{l|}{29.40}         & \multicolumn{1}{l|}{30.23}      & \multicolumn{1}{l|}{29.13}      & \multicolumn{1}{l|}{29.42}       & \multicolumn{1}{l|}{32.45}     & \multicolumn{1}{l|}{30.01}        & \multicolumn{1}{l|}{30.22}     & \multicolumn{1}{l|}{30.11}    & \multicolumn{1}{l!{\vrule width1pt}}{30.12}       & \multicolumn{1}{l!{\vrule width1pt}}{30.43}                 \\ \hline

\multirowthead{1}{\textbf{Ours}}        & \multicolumn{1}{l|}{29.82}      & \multicolumn{1}{l|}{\textbf{33.82}}      & \multicolumn{1}{l|}{\textbf{31.47}}        & \multicolumn{1}{l|}{\textbf{30.10}}         & \multicolumn{1}{l|}{\textbf{30.67}}      & \multicolumn{1}{l|}{\textbf{29.50}}     & \multicolumn{1}{l|}{\textbf{29.87}}       & \multicolumn{1}{l|}{\textbf{33.09}}     & \multicolumn{1}{l|}{\textbf{31.32}}        & \multicolumn{1}{l|}{\textbf{30.86}}     & \multicolumn{1}{l|}{\textbf{30.64}}    & \multicolumn{1}{l!{\vrule width1pt}}{\textbf{30.77}}       & \multicolumn{1}{l!{\vrule width1pt}}{\textbf{30.99}}                 \\ \Xhline{1pt}

\multirowthead{1}{Noise Lv} & \multicolumn{13}{c!{\vrule width1pt}}{$\sigma=50$}                                                                                                                                                                                                                                                                                                                                                             \\ \Xhline{1pt}
\multirowthead{1}{\cite{BM3D}}        & \multicolumn{1}{l|}{26.13}      & \multicolumn{1}{l|}{29.69}      & \multicolumn{1}{l|}{26.68}        & \multicolumn{1}{l|}{25.04}         & \multicolumn{1}{l|}{25.82}      & \multicolumn{1}{l|}{25.10}      & \multicolumn{1}{l|}{25.90}       & \multicolumn{1}{l|}{29.05}     & \multicolumn{1}{l|}{27.23}        & \multicolumn{1}{l|}{26.78}     & \multicolumn{1}{l|}{26.81}    & \multicolumn{1}{l!{\vrule width1pt}}{26.46}       & \multicolumn{1}{l!{\vrule width1pt}}{26.73}                 \\ \hline
\multirowthead{1}{\cite{WNNM}}        & \multicolumn{1}{l|}{26.42}      & \multicolumn{1}{l|}{30.33}      & \multicolumn{1}{l|}{26.91}        & \multicolumn{1}{l|}{25.43}         & \multicolumn{1}{l|}{26.32}      & \multicolumn{1}{l|}{25.42}      & \multicolumn{1}{l|}{26.09}       & \multicolumn{1}{l|}{29.25}     & \multicolumn{1}{l|}{\textbf{27.79}}        & \multicolumn{1}{l|}{26.97}     & \multicolumn{1}{l|}{26.94}    & \multicolumn{1}{l!{\vrule width1pt}}{26.64}       & \multicolumn{1}{l!{\vrule width1pt}}{27.04}                 \\ \hline
\multirowthead{1}{\cite{Zoran:ICCV11}}        & \multicolumn{1}{l|}{26.02}      & \multicolumn{1}{l|}{28.76}      & \multicolumn{1}{l|}{26.63}        & \multicolumn{1}{l|}{25.04}         & \multicolumn{1}{l|}{25.78}      & \multicolumn{1}{l|}{25.24}      & \multicolumn{1}{l|}{25.84}       & \multicolumn{1}{l|}{28.43}     & \multicolumn{1}{l|}{24.82}        & \multicolumn{1}{l|}{26.65}     & \multicolumn{1}{l|}{26.72}    & \multicolumn{1}{l!{\vrule width1pt}}{26.24}       & \multicolumn{1}{l!{\vrule width1pt}}{26.35}                 \\ \hline
\multirowthead{1}{\cite{TNRD}}        & \multicolumn{1}{l|}{26.62}      & \multicolumn{1}{l|}{29.48}      & \multicolumn{1}{l|}{27.10}        & \multicolumn{1}{l|}{25.42}         & \multicolumn{1}{l|}{26.31}      & \multicolumn{1}{l|}{25.59}      & \multicolumn{1}{l|}{26.16}       & \multicolumn{1}{l|}{28.93}     & \multicolumn{1}{l|}{25.70}        & \multicolumn{1}{l|}{26.94}     & \multicolumn{1}{l|}{26.98}    & \multicolumn{1}{l!{\vrule width1pt}}{26.50}       & \multicolumn{1}{l!{\vrule width1pt}}{26.81}                 \\ \hline
\multirowthead{1}{\cite{Zhang:TIP17}}       & \multicolumn{1}{l|}{\textbf{27.00}}      & \multicolumn{1}{l|}{30.02}      & \multicolumn{1}{l|}{27.29}        & \multicolumn{1}{l|}{25.70}         & \multicolumn{1}{l|}{26.77}      & \multicolumn{1}{l|}{25.87}      & \multicolumn{1}{l|}{26.48}       & \multicolumn{1}{l|}{29.37}     & \multicolumn{1}{l|}{26.23}        & \multicolumn{1}{l|}{27.19}     & \multicolumn{1}{l|}{27.24}    & \multicolumn{1}{l!{\vrule width1pt}}{26.90}       & \multicolumn{1}{l!{\vrule width1pt}}{27.17}                 \\ \hline

\multirowthead{1}{\textbf{Ours}}        & \multicolumn{1}{l|}{26.90}      & \multicolumn{1}{l|}{\textbf{30.50}}      & \multicolumn{1}{l|}{\textbf{27.89}}        & \multicolumn{1}{l|}{\textbf{26.46}}         & \multicolumn{1}{l|}{\textbf{27.37}}      & \multicolumn{1}{l|}{\textbf{26.35}}      & \multicolumn{1}{l|}{\textbf{26.96}}       & \multicolumn{1}{l|}{\textbf{29.87}}     & \multicolumn{1}{l|}{27.17}        & \multicolumn{1}{l|}{\textbf{27.74}}     & \multicolumn{1}{l|}{\textbf{27.67} }   & \multicolumn{1}{l!{\vrule width1pt}}{\textbf{27.41}}       & \multicolumn{1}{l!{\vrule width1pt}}{\textbf{27.69}}                 \\ \Xhline{1pt}
\end{tabular}
\label{den-set12}
\end{table*}

\begin{figure*}[!tbh]
\renewcommand{\arraystretch}{0.4}
\centering

\subfigure[]{
    \includegraphics[width=0.28\textwidth]{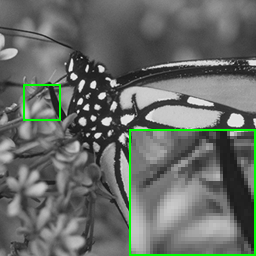}
    }\hspace{-0.8em}
\subfigure[]{
    \includegraphics[width=0.28\textwidth]{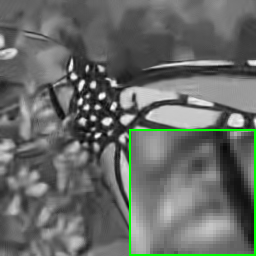}
    }\hspace{-0.8em}
\subfigure[]{
    \includegraphics[width=0.28\textwidth]{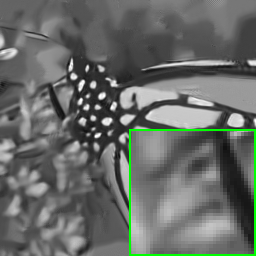}
    }\hspace{-0.8em}
\subfigure[]{
    \includegraphics[width=0.28\textwidth]{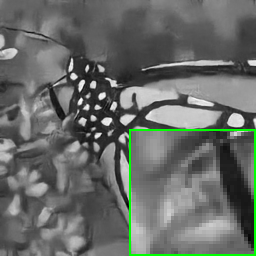}
    }\hspace{-0.8em}
\subfigure[]{
    \includegraphics[width=0.28\textwidth]{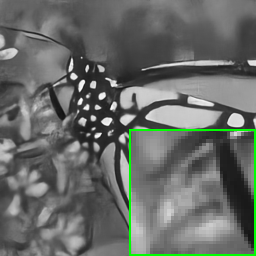}
    }\hspace{-0.8em}
\subfigure[]{
    \includegraphics[width=0.28\textwidth]{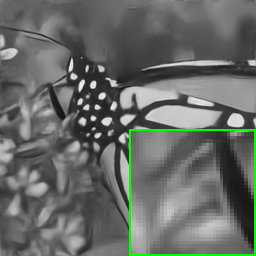}
    }\hspace{-0.8em}

     \caption{Denoising results of noise level of 50. (a) Parts of the original images of ``monarch'' in Set12; (b) BM3D\cite{BM3D}(PSNR=25.82dB); (c) WNNM\cite{WNNM}(PSNR=26.32dB); (d) TNRD\cite{TNRD}(PSNR=26.31dB);(e) DnCNN\cite{Zhang:TIP17}(PSNR=26.77dB); (f) \bf{Proposed method(PSNR=27.37dB)}}
\end{figure*}

\begin{table*}[tbh]
\centering
\caption{Results of different deblurring methods}
\begin{tabular}{!{\vrule width1pt}c!{\vrule width1pt}c!{\vrule width1pt}c|c|c|c|c|c|c|c|c|c|c!{\vrule width1pt}}
\Xhline{1pt}
Methods  &      $\sigma$       & Butt. & Pepp. & Parr. & star. & Barb. & Boats & C.Man & House  & Leaves & Lena     & Avg \\ \Xhline{1pt}

\multicolumn{13}{!{\vrule width1pt}c!{\vrule width1pt}}{Kenel 1 (19*19)}                                                                                                 \\ \Xhline{1pt}
\cite{Zoran:ICCV11}     & \multirow{3}{*}{2.6} &      26.23     &   27.40    &    33.78   &  29.79     &    29.78    &  30.15    &        30.24 &     31.73     &    25.84     &    31.37    &     29.63    \\ \cline{1-1} \cline{3-13}
\cite{Zhang:CVPR17}    &                       &    32.23       &    32.00   &   34.48    &   32.26    &       32.38 &   33.05   &   31.50      &    34.89      &     33.29    &  33.54      &    32.96     \\ \cline{1-1} \cline{3-13}

\textbf{Ours}     &                       &     \textbf{32.58}      &   \textbf{32.36}    &  \textbf{34.63}     &   \textbf{32.54}    &       \textbf{32.52} &  \textbf{33.27}    &     \textbf{31.83}    &     \textbf{35.03}     &    \textbf{33.30}     &  \textbf{33.66}      &    \textbf{ 33.17 }    \\ \Xhline{1pt}
\cite{Zoran:ICCV11}     & \multirow{3}{*}{7.7} &     24.27      &   26.15    &   30.01    &  26.81     &    26.95    &  27.72    &        27.37 &     29.89     &     23.81    &   28.69     &    27.17     \\ \cline{1-1} \cline{3-13}
\cite{Zhang:CVPR17}    &                       &   28.51       &   28.88   &  \textbf{31.07 }   &  27.86     &       \textbf{28.18} &   \textbf{29.13}   &     28.11    &   \textbf{32.03}       &   \textbf{28.42}      &  \textbf{29.52}     &    29.17     \\ \cline{1-1} \cline{3-13}

\textbf{Ours}     &                       &     \textbf{28.53}     &   \textbf{28.88}   &   31.06    &    \textbf{27.93}    &  28.17 & 29.11    &       \textbf{28.14} &   31.94   &    28.40     &    29.49    &   \textbf{29.17} \\ \Xhline{1pt}
\multicolumn{13}{!{\vrule width1pt}c!{\vrule width1pt}}{Kenel 2 (17*17)}                                                                                                 \\ \Xhline{1pt}
\cite{Zoran:ICCV11}     & \multirow{3}{*}{2.6} &     26.48      &   27.37    &   33.88    &   29.56    &   28.29     &   29.61   &     29.66    &   32.97       &    25.69     &   30.67     &     29.42    \\ \cline{1-1} \cline{3-13}
\cite{Zhang:CVPR17}    &                       &   31.97       &   31.89    &    34.46   &   32.18    &       32.00 & 33.06    &   31.29     &     34.82     &    32.96     &   33.35     &    32.80     \\ \cline{1-1} \cline{3-13}

\textbf{Ours}     &                      &        \textbf{32.22}    &    \textbf{32.16}        &    \textbf{34.57}   &  \textbf{32.36}     &   \textbf{32.06}    &      \textbf{33.17}   &    \textbf{31.52}   &    \textbf{34.99}     &    \textbf{32.96}      &    \textbf{33.41}     &   \textbf{32.94}      \\ \Xhline{1pt}
\cite{Zoran:ICCV11}     & \multirow{3}{*}{7.7} &     23.85      &   26.04    &   29.99    &    26.78   &   25.47     &   27.46   &    26.58     &     30.49     &    23.42     &    28.20    &     26.83    \\ \cline{1-1} \cline{3-13}
\cite{Zhang:CVPR17}    &                       &      \textbf{28.21}     &   \textbf{28.71}    &   \textbf{30.68}    &   27.67    &       27.37 &   28.95   &   27.70      &    \textbf{31.95}      &   27.92     &    \textbf{29.27}    &    28.84     \\ \cline{1-1} \cline{3-13}

\textbf{Ours}     &                       &      28.20     &   28.71 &    30.64   &   \textbf{27.73}    &   \textbf{27.47}    &    \textbf{28.97}   &    \textbf{27.71}  &    31.86     &     \textbf{27.94}      &    29.21          &         \textbf{28.84}\\ \Xhline{1pt}
\end{tabular}
\label{deblur-set10}
\end{table*}

\begin{figure*}[!tbh]
\renewcommand{\arraystretch}{0.4}
\centering
\subfigure[]{
    \includegraphics[width=0.22\textwidth]{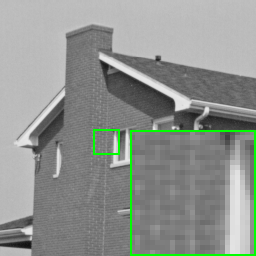}
    }\hspace{-0.8em}
\subfigure[]{
    \includegraphics[width=0.22\textwidth]{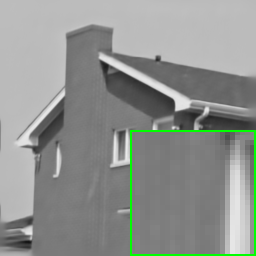}
    }\hspace{-0.8em}
\subfigure[]{
    \includegraphics[width=0.22\textwidth]{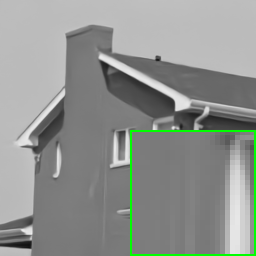}
    }\hspace{-0.8em}
\subfigure[]{
    \includegraphics[width=0.22\textwidth]{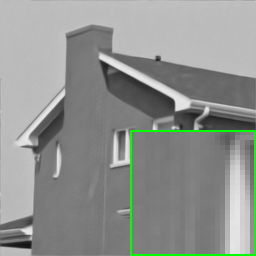}
    }\hspace{-0.8em}

     \caption{Deblurring results at the noise level of 2.55, kernel 1. (a) Parts of the original images of ``house'' in Set10; (b) EPLL\cite{Zoran:ICCV11}(PSNR=32.13dB); (c) IR-CNN\cite{Zhang:CVPR17}(PSNR=34.87dB); (d)\textbf{Proposed method(PSNR=35.53dB)}}
\label{fig:deblur1}
\end{figure*}

\begin{figure*}[!tbh]
\renewcommand{\arraystretch}{0.4}
\centering
\subfigure[]{
    \includegraphics[width=0.22\textwidth]{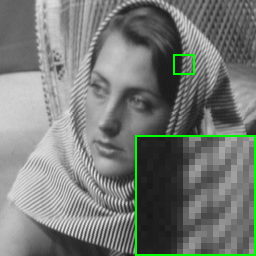}
    }\hspace{-0.8em}
\subfigure[]{
    \includegraphics[width=0.22\textwidth]{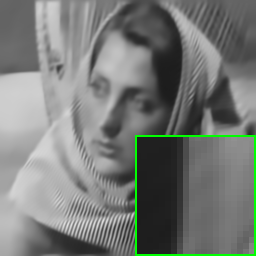}
    }\hspace{-0.8em}
\subfigure[]{
    \includegraphics[width=0.22\textwidth]{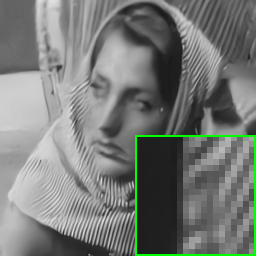}
    }\hspace{-0.8em}
\subfigure[]{
    \includegraphics[width=0.22\textwidth]{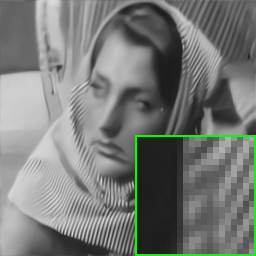}
    }\hspace{-0.8em}

    \caption{Deblurring results at the noise level of 7.65, kernel 2. (a) Parts of the original images of ``barbara'' in Set10; (b) EPLL\cite{Zoran:ICCV11}(PSNR=25.70dB); (c) IR-CNN\cite{Zhang:CVPR17}(PSNR=27.38dB); (d)\bf{Proposed method(PSNR=27.73dB)}}
\label{fig:deblur2}
\end{figure*}

\subsection{Image super-resolution}

\begin{figure*}[!tbh]
\renewcommand{\arraystretch}{0.4}
\centering
\subfigure[]{
    \includegraphics[width=0.28\textwidth]{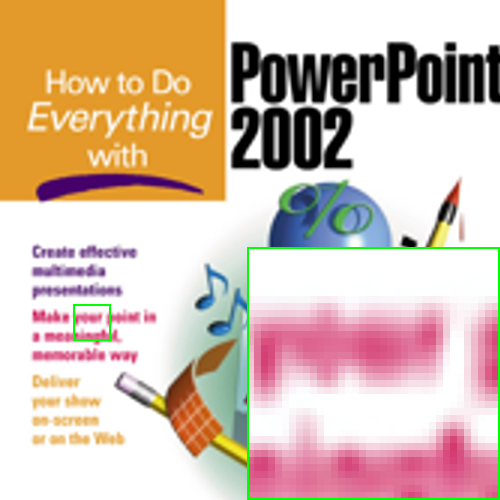}
    }\hspace{-0.8em}
\subfigure[]{
    \includegraphics[width=0.28\textwidth]{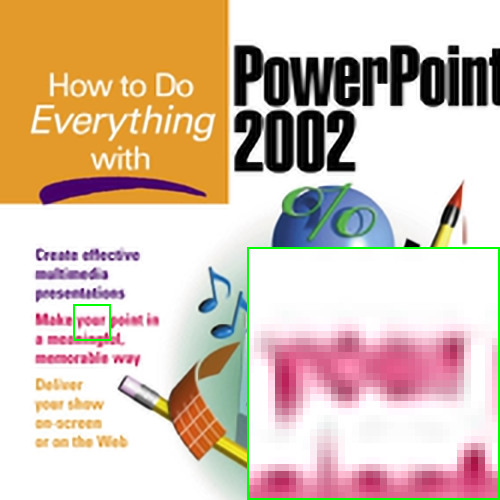}
    }\hspace{-0.8em}
\subfigure[]{
    \includegraphics[width=0.28\textwidth]{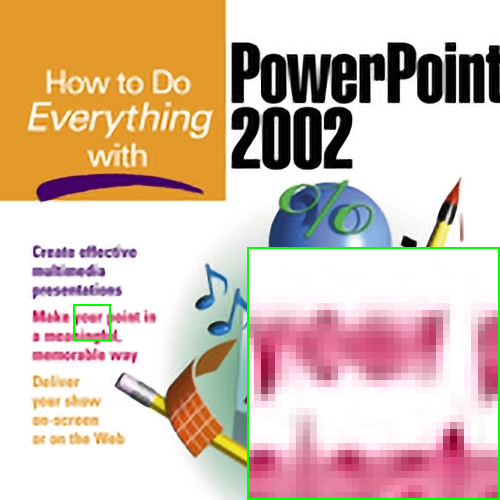}
    }\hspace{-0.8em}
\subfigure[]{
    \includegraphics[width=0.28\textwidth]{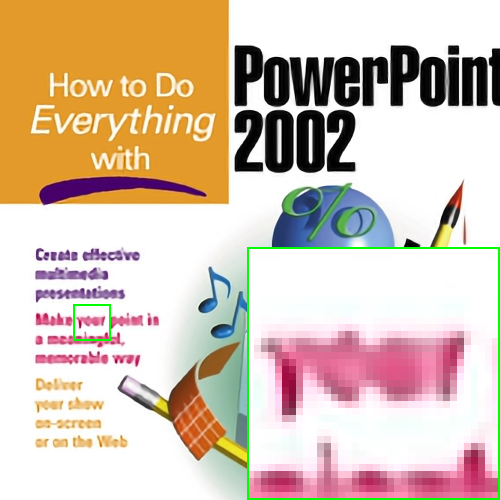}
    }\hspace{-0.8em}
\subfigure[]{
    \includegraphics[width=0.28\textwidth]{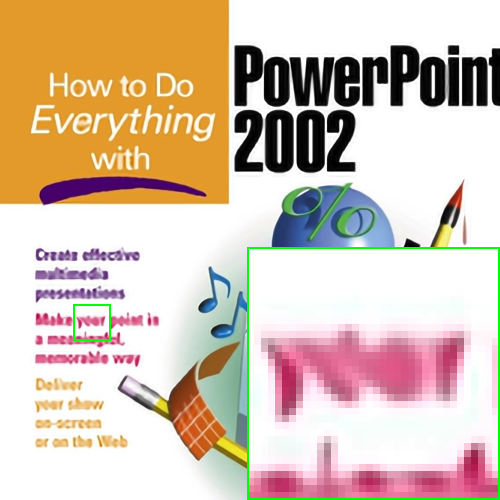}
    }\hspace{-0.8em}
\subfigure[]{
    \includegraphics[width=0.28\textwidth]{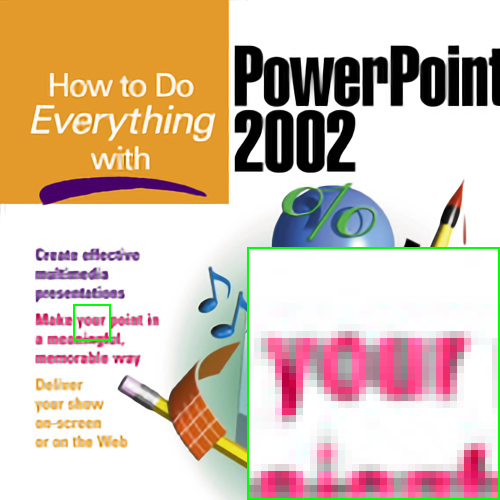}
    }\hspace{-0.8em}
    \\
     \caption{SR results of scaling factor of 3. (a) Parts of the original images of ``ppt3'' in Set14; (b) NCSR\cite{dong2013nonlocally}(PSNR=25.66dB); (c) SRCNN\cite{SRCNN}(PSNR=27.04dB); (d) VDSR\cite{VDSR}(PSNR=27.86dB);(e) DRCN\cite{DRCN}(PSNR=27.73dB); (f) \bf{Proposed method(PSNR=28.16dB)}}
\label{fig:SR1}
\end{figure*}

\begin{figure*}[!tbh]
\renewcommand{\arraystretch}{0.4}
\centering
    \subfigure[]{
    \includegraphics[width=0.28\textwidth]{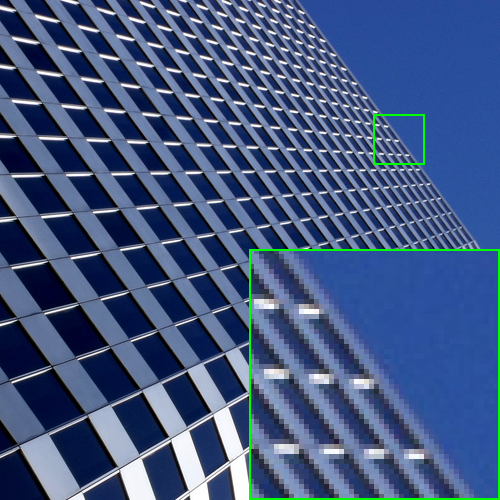}
    }\hspace{-0.8em}
\subfigure[]{
    \includegraphics[width=0.28\textwidth]{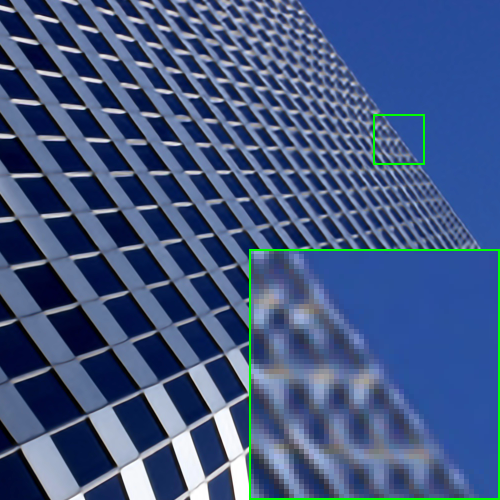}
    }\hspace{-0.8em}
\subfigure[]{
    \includegraphics[width=0.28\textwidth]{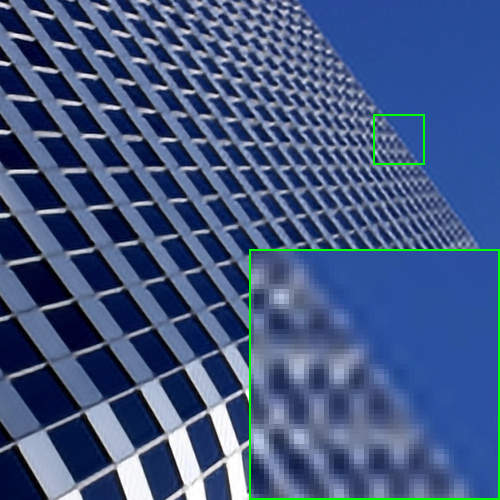}
    }\hspace{-0.8em}
\subfigure[]{
    \includegraphics[width=0.28\textwidth]{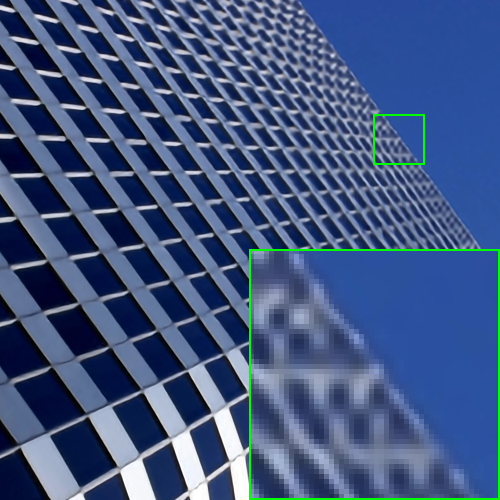}
    }\hspace{-0.8em}
\subfigure[]{
    \includegraphics[width=0.28\textwidth]{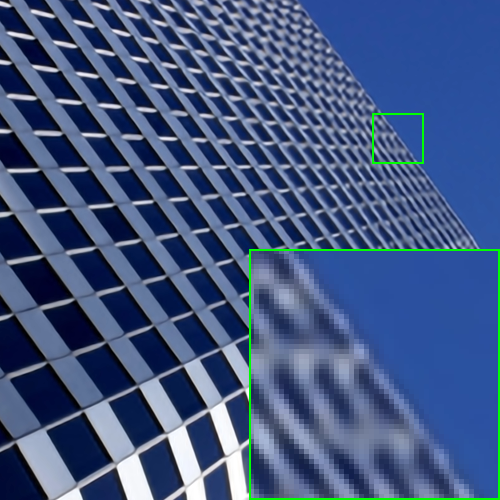}
    }\hspace{-0.8em}
\subfigure[]{
    \includegraphics[width=0.28\textwidth]{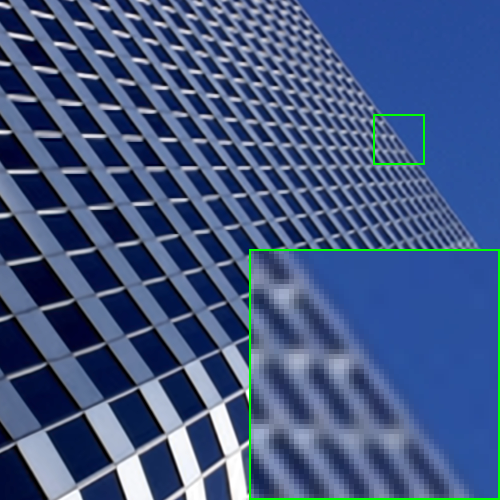}
    }\hspace{-0.8em}
    \\
    \caption{SR results of scaling factor of 4. (a) Parts of the original images of ``img005'' in Urban100 dataset; (b) NCSR\cite{dong2013nonlocally}(PSNR=26.44dB); (c) SRCNN\cite{SRCNN}(PSNR=25.50dB); (d) VDSR\cite{VDSR}(PSNR=26.70dB);(e) DRCN\cite{DRCN}(PSNR=26.82dB); (f)\bf{ Proposed method(PSNR=27.01dB)}}

\label{fig:SR2}
\end{figure*}

With augmentation, $600, 000$ pairs of LR/HR image patches were extracted from the pair of LR/HR training images. The LR patch is of size $40\times 40$ and the HR patch is sized by $40s\times 40s$; we have trained a separate network for each scaling factor $s$($s=2,3,4$). The commonly used datasets, including Set5, Set14, the BSD100, and the Urban 100 dataset \cite{kim2016accurate} containing 100 high-quality images were used in our experiments. %Tables \ref{SR-Var-bicubic} and \ref{SR-Var-Gaussian} show the average PSNR/SSIM results of the variants of the proposed method on several commonly used image sets. Similar to image denoising and deblurring, we have observed that SASC-CNN-SS method consistently outperform ASC and SASC-SS.
We have compared the proposed method against several leading deep learning based image SR methods including SRCNN \cite{SRCNN}, VDSR \cite{kim2016accurate} and DRCN\cite{DRCN}, and denoising-based SR methods (i.e. TNRD\cite{TNRD}). For fair comparisons, the results ofall benchamrk methods are either directly cited from their papers or generated by the codes released by the authors. The PSNR results of these competing methods for the bicubic case are shown in Tables \ref{SR-set5}-\ref{SR-others}, from which one can see that the proposed method outperforms other competing methods. Portions of reconstructed HR images by different methods are shown in Figs. \ref{fig:SR1} and \ref{fig:SR2}. It can be seen that the proposed method can more faithfully restore fine text details, while other methods including VDSR \cite{kim2016accurate} fail to deliver the same.

%\clearpage
\begin{table*}[tbh]
\centering
\caption{Results of proposed super-resolution method in Set5}
\setlength{\tabcolsep}{3.5pt}
\begin{tabular}{!{\vrule width1pt}c|c|c|c|c|c|c!{\vrule width1pt}}
\Xhline{1pt}
Images    & Scale                & TNRD\cite{TNRD} & SRCNN\cite{SRCNN} & VDSR\cite{VDSR} & DRCN\cite{DRCN} & Ours \\ \Xhline{1pt}
Baby      & \multirow{6}{*}{2}         &   38.53   &   38.54    &  38.75    &  38.80     &\textbf{38.83}      \\ \cline{1-1} \cline{3-7}
Bird      &                            &   41.31   &  40.91     &   42.42   &    42.68   &  \textbf{42.70}   \\ \cline{1-1} \cline{3-7}
Butterfly &                            &  33.17    &   32.75    &   34.49   &    34.56   &   \textbf{34.72}  \\ \cline{1-1} \cline{3-7}
Head      &                            &  35.75    &  35.72     &   35.93   &   35.95    &   \textbf{35.96}   \\ \cline{1-1} \cline{3-7}
Woman     &                            &  35.50    &   35.37    &   36.05   &     36.15  &   \textbf{36.33}  \\ \Xcline{1-1}{1.0pt}
\Xcline{3-7}  {1pt}
\textbf{Average}                       &        &  36.85    &   36.66    &   37.53   &   37.63    &  \textbf{37.71}    \\ \Xhline{1pt}
Baby      & \multirow{6}{*}{3} &          35.28   &   35.25    &   35.38   &   35.50    & \textbf{35.56}      \\ \cline{1-1} \cline{3-7}
Bird      &                    &           36.09   &   35.48    &   36.66   &    37.05   &  \textbf{37.20}   \\ \cline{1-1} \cline{3-7}
Butterfly &                    &          28.92    &  27.95     &  29.96   &     30.03  &   \textbf{30.23}   \\ \cline{1-1} \cline{3-7}
Head      &                    &          33.75    &   33.71    &    33.96  &     34.00  &   \textbf{34.01}   \\ \cline{1-1}\cline{3-7}
Woman     &                    &          31.79    &    31.37   &   32.36   &    32.53   &  \textbf{32.63}   \\ \Xcline{1-1}{1.0pt}
\Xcline{3-7} {1pt}
\textbf{Average}   &                             &  33.17    &   32.75    &  33.66    &    33.82   &  \textbf{33.93}    \\ \Xhline{1pt}
Baby      & \multirow{6}{*}{4}     &  31.30    &   33.13    &  33.41    &    33.51   &    \textbf{33.61} \\ \cline{1-1} \cline{3-7}
Bird      &                            &   32.99   &   32.52    &    33.54  &    33.78   & \textbf{33.93}     \\ \cline{1-1} \cline{3-7}
Butterfly &                            &  26.22    &   25.46    &    27.28  &     27.47  &  \textbf{27.56}   \\ \cline{1-1} \cline{3-7}
Head      &                            &  32.51    &   32.44    &   32.70   &    32.82   &   \textbf{32.82}   \\ \cline{1-1} \cline{3-7}
Woman     &                            &  29.20    &   28.89    &    29.81  &     30.09  &   \textbf{30.20}   \\ \Xcline{1-1}{1pt}
\Xcline{3-7} {1.0pt}
\textbf{Average}   &                             &  30.85    &    30.48   &  31.35    &    31.53   &  \textbf{31.62}    \\ \Xhline{1pt}

\end{tabular}
\label{SR-set5}
\end{table*}

%\clearpage
\begin{table*}[tbh]
\centering
\caption{Results of proposed super-resolution method in Set14, BSD100 and Urban100}
\begin{tabular}{!{\vrule width1pt}c!{\vrule width1pt}c|c|c|c|c|c|c|c|c|c|c!{\vrule width1pt}}
\Xhline{1pt}
\multirow{2}{*}{Dataset}  & \multirow{2}{*}{Scale}    & \multicolumn{2}{c|}{TNRD\cite{TNRD}}& \multicolumn{2}{c|}{SRCNN\cite{SRCNN}} & \multicolumn{2}{c|}{VDSR\cite{VDSR}} & \multicolumn{2}{c|}{DRCN\cite{DRCN}} & \multicolumn{2}{c!{\vrule width1pt}}{Ours} \\ \cline{3-12}
                          &                                                         & PSNR         & SSIM        & PSNR        & SSIM        & PSNR        & SSIM       & PSNR         & SSIM        & PSNR        & SSIM        \\ \Xhline{1pt}

\multirow{3}{*}{Set14}    & 2                                   &      32.54  &    0.907         &      32.42         &   0.906   &       33.03        &       0.912 &     33.04     &    0.912    &      \textbf{33.20}  &    \textbf{0.914}\\ \cline{2-2} \cline{2-12}
                          & 3                                                             &      29.46  &    0.823   &       29.28        &   0.821          &   29.77    &            0.831 &    29.76    &                   0.831 &    \textbf{29.96}   &    \textbf{0.835 }   \\ \cline{2-2} \cline{2-12}
                          & 4                                                             &     27.68   &     0.756  &      27.49         &     0.750        &   28.01          &            0.767 &        28.02        &     0.767    &   \textbf{ 28.15 } &  \textbf{0.770}   \\ \cline{2-12}  \Xhline{1pt}
\multirow{3}{*}{BSD100}   & 2                                    &      31.40   &   0.888     &     31.36         &    0.888         &     31.90         &            0.896 &   31.85   & 0.894  &     \textbf{ 31.94 }  &   \textbf{0.896}\\ \cline{2-2} \cline{2-12}
                          & 3                                                             &     28.50         &     0.788        &      28.41        &    0.786         &              28.80 &       0.796      &   28.80     &      0.795 &     \textbf{28.88}     & \textbf{0.799} \\ \cline{2-2} \cline{2-12}
                          & 4                                                            &      27.00   &    0.714    &      26.90       &    0.710         &    27.23          &            0.723 &         27.08    &            0.709 &    \textbf{ 27.33 }   & \textbf{0.726} \\ \cline{2-12} \Xhline{1pt}
\multirow{3}{*}{Urban100} & 2                                   &      29.70    &     0.899       &     29.50          &     0.895        &             30.76 &      0.914       &   30.75      &        0.913      &   \textbf{ 30.97 }   &     \textbf{0.915}\\ \cline{2-2} \cline{3-12}
                          & 3                                                             &      26.44    &     0.807        &     26.24         &    0.799         &             27.15 &        0.828     &    27.08    &             0.824 &    \textbf{27.33}     &     \textbf{0.831}\\ \cline{2-2} \cline{3-12}
                          & 4                                                             &     24.62   &     0.729     &     24.52         &    0.722         &             25.14 &        0.751     &       24.94      &     0.735 &     \textbf{ 25.33 }  &   \textbf{0.756} \\ \cline{2-12} \Xhline{1pt}
\end{tabular}
\label{SR-others}
\end{table*}

\section{Conclusion}

In this paper, we propose a structured analysis sparse coding (SASC) based network for image restoration and show that the structured sparse prior learned from both large-scale training dataset and the input degraded image can significantly improve the sparsity-based performance. Furthermore, we propose a network implementation of the SASC for image restoration for efficiency and better performance. Experimental results show that the proposed method performs comparably to and often even better than the current state-of-the-art restoration methods.

%\clearpage

\ifCLASSOPTIONcaptionsoff
  \newpage
\fi

\bibliographystyle{IEEEtran}
\bibliography{SASC}

\end{document}